\documentclass[10pt,conference]{IEEEtran} 
\IEEEoverridecommandlockouts
\usepackage{cite}
\usepackage{amsmath,amssymb,amsfonts}
\usepackage{algorithmic}
\usepackage{graphicx}
\usepackage{textcomp}
\usepackage{xcolor}
\usepackage{booktabs}
\usepackage{comment}
\usepackage{tabularx}
\usepackage{enumitem}
\newcolumntype{Y}{>{\raggedright\arraybackslash}X}

\def\BibTeX{{\rm B\kern-.05em{\sc i\kern-.025em b}\kern-.08em
    T\kern-.1667em\lower.7ex\hbox{E}\kern-.125emX}}
\begin{document}

\title{Failure-Aware Enhancements for Large Language Model (LLM) Code Generation: An Empirical Study on Decision Framework\\
% {\footnotesize \textsuperscript{*}Note: Sub-titles are not captured in Xplore and
% should not be used}
% \thanks{Identify applicable funding agency here. If none, delete this.}
}
% When Prompting Fails: A Failure-Aware Decision Framework for Enhancing LLM-Based Code Generation

% \author{\IEEEauthorblockN{Jianru Shen, Zedong Peng\textsuperscript{*} and Lucy Owen}
% \IEEEauthorblockA{
% \textit{University of Montana, MT, USA}\\
% jianru.shen@umconnect.umt.edu, zedong.peng@mso.umt.edu, lucy.owen@mso.umt.edu}
% }

\author{%
\IEEEauthorblockN{Jianru Shen\textsuperscript{*}, Zedong Peng\textsuperscript{*}, Lucy Owen}
\IEEEauthorblockA{\textit{University of Montana, Missoula, MT, USA}\\
jianru.shen@umconnect.umt.edu, zedong.peng@mso.umt.edu, lucy.owen@mso.umt.edu}
\thanks{\textsuperscript{*}Corresponding authors}
}
% \author{
%     \IEEEauthorblockN{Jianru Shen, Zedong Peng\IEEEauthorrefmark{1} and Lucy Owen}
%     \IEEEauthorblockA{
%         \textit{University of Montana, MT, USA}\\
%         \{jianru.shen, zedong.peng, lucy.owen\}@mso.umt.edu
%     }
%     \thanks{\IEEEauthorrefmark{1}Corresponding author: zedong.peng@mso.umt.edu}
% }
% \and
% \IEEEauthorblockN{2\textsuperscript{nd} Given Name Surname}
% \IEEEauthorblockA{\textit{dept. name of organization (of Aff.)} \\
% \textit{name of organization (of Aff.)}\\
% City, Country \\
% email address or ORCID}
% \and
% \IEEEauthorblockN{3\textsuperscript{rd} Given Name Surname}
% \IEEEauthorblockA{\textit{dept. name of organization (of Aff.)} \\
% \textit{name of organization (of Aff.)}\\
% City, Country \\
% email address or ORCID}
% \and
% \IEEEauthorblockN{4\textsuperscript{th} Given Name Surname}
% \IEEEauthorblockA{\textit{dept. name of organization (of Aff.)} \\
% \textit{name of organization (of Aff.)}\\
% City, Country \\
% email address or ORCID}
% \and
% \IEEEauthorblockN{5\textsuperscript{th} Given Name Surname}
% \IEEEauthorblockA{\textit{dept. name of organization (of Aff.)} \\
% \textit{name of organization (of Aff.)}\\
% City, Country \\
% email address or ORCID}
% \and
% \IEEEauthorblockN{6\textsuperscript{th} Given Name Surname}
% \IEEEauthorblockA{\textit{dept. name of organization (of Aff.)} \\
% \textit{name of organization (of Aff.)}\\
% City, Country \\
% email address or ORCID}
% }

\maketitle

\begin{abstract}

Large language models (LLMs) show promise for automating software development by translating requirements into code. However, even advanced prompting workflows like progressive prompting often leave some requirements unmet. Although methods such as self-critique, multi-model collaboration, and retrieval-augmented generation (RAG) have been proposed to address these gaps, developers lack clear guidance on when to use each. In an empirical study of 25 GitHub projects, we found that progressive prompting achieves 96.9\% average task completion, 
significantly outperforming direct prompting (80.5\%, Cohen's d=1.63, 
\textbf{\boldmath $p<0.001$}) but still leaving 8 projects incomplete. For 6 of the most representative projects, we evaluated each enhancement strategy across 4 failure types. Our results reveal that method effectiveness depends critically on 
failure characteristics: Self-Critique succeeds on code-reviewable 
logic errors but fails completely on external service integration (0\% improvement), while RAG achieves highest completion across all failure types with superior efficiency. Based on these findings, we propose a decision framework that maps each failure pattern to the most suitable enhancement method, giving practitioners practical, data-driven guidance instead of trial-and-error.

% Large language models (LLMs) show promise for automated code generation, 
% yet even advanced prompting strategies such as progressive prompting fail to 
% achieve complete requirement coverage. 
% While various enhancement methods---including self-critique, multi-model 
% collaboration, and retrieval-augmented generation (RAG)---have been 
% proposed to address these failures, practitioners lack systematic guidance 
% on when to apply each method. 
% We conduct an empirical study across 25 GitHub projects, finding that 
% progressive prompting achieves 96.9\% average task completion, 
% significantly outperforming direct prompting (80.5\%, Cohen's d=1.63, 
% \textbf{\boldmath $p<0.001$}) but still leaving 8 projects incomplete. 
% We design and evaluate three enhancement methods on incomplete challenge projects 
% representing diverse failure types. 
% Our results reveal that method effectiveness depends critically on 
% failure characteristics: Self-Critique succeeds on code-reviewable 
% logic errors but fails completely on external service integration (0\% improvement), while RAG achieves highest completion across all failure types with superior efficiency. 
% Based on these findings, we propose a decision framework that maps each failure pattern to the most suitable enhancement method, giving practitioners practical, data-driven guidance instead of trial-and-error.

% We provide a decision framework mapping failure types to appropriate 
% enhancement methods, enabling practitioners to select strategies 
% based on project characteristics rather than trial-and-error.
\end{abstract}

\begin{comment}
an alternative version as a reference:
Progressive prompting significantly improves LLM-based code generation over 
direct prompting (96.9% vs 80.5% task completion, Cohen's d=1.63, p<0.001), 
yet failures persist in 8 of 25 evaluated projects. While enhancement methods 
exist—self-critique, multi-model collaboration, and retrieval-augmented 
generation (RAG)—practitioners lack systematic guidance on method selection. 
We conduct an empirical study evaluating these three methods on six challenge 
projects representing five failure types: CRUD operations, external service 
integration, business logic, domain-specific knowledge, and infrastructure 
configuration. Our results reveal that method effectiveness depends on failure 
characteristics. Self-Critique achieves 100% completion on code-reviewable 
failures (CRUD, logic errors) in 138 minutes average but fails completely 
on external knowledge failures (0% improvement). RAG succeeds across all 
failure types with superior efficiency (31.5 min/pp vs 40.6 for Self-Critique, 
47.4 for Multi-Model). We contribute a decision framework mapping failure 
characteristics to appropriate enhancement methods, demonstrating that 
code-reviewable failures benefit from Self-Critique while external service 
integration, domain-specific, and infrastructure failures require RAG.
\end{comment}


\section{Introduction}
% Large language models (LLMs) have transformed automated code generation, with tools like GitHub Copilot and ChatGPT demonstrating impressive capabilities across diverse programming tasks. Progressive prompting---a multi-phase strategy that decomposes code generation into sequential steps (requirements analysis $\rightarrow$ architectural design $\rightarrow$ test specification $\rightarrow$ implementation)---has emerged as the current best practice, significantly outperforming single-step direct prompting approaches.

Imagine a software engineer using an AI pair programmer to implement a new web service. Initially, the large language model (LLM) assistant (e.g., GitHub Copilot or ChatGPT) produces promising code for basic CRUD functionalities. The engineer follows a progressive prompting workflow: first prompting for requirements analysis, then architectural design, test specification, and finally code implementation, which is a strategy that generally yields better results than one-shot code generation \cite{wei2024requirements}. Yet, as the project progresses, the assistant struggles with certain tasks. For example, when integrating an external payment API, the generated code omits essential steps and fails to fully meet the requirement. The developer is left wondering: should they prompt the LLM to critique and refine its output, fetch relevant API documentation for the model, or involve a different model altogether? This scenario highlights a real-world challenge: even the best prompting practices can systematically fail on complex requirements, and it is unclear which remedial strategy to apply for a given failure.

% Progressive prompting still misses some requirements. In a baseline study of 25 GitHub projects, it significantly outperformed direct prompting in coverage, but some projects remained incompletely implemented. In our study, 8 out of 25 projects (32\%) remained incomplete. Prior work has proposed a variety of enhancement strategies to close these gaps: such as iterative self-critique \cite{yao2025mcqg}, multi-model collaboration \cite{dong2025effiqa,wen2025ai}, and retrieval-augmented generation (RAG) \cite{zhao2024retrieval}, but these techniques have not been systematically organized or compared by failure mode. In this paper, we synthesize the literature into three representative enhancement families and instantiate each with a concrete method. The three enhancement methods we study are not merely abstract categories: in this work, they are concretely instantiated, tuned, and integrated by us into practical pipelines for LLM-based development. At the end, we provide structured guidance on \textbf{when} each family helps, \textbf{which} failure patterns it mitigates, and \textbf{where} gaps remain. We frame our analysis around three research questions

Progressive prompting still misses some requirements. In our baseline study of 25 GitHub projects, it significantly outperformed direct prompting in coverage, but some projects remained incompletely implemented. We found that 8 out of 25 projects (32\%) remained incomplete. Prior work has proposed a variety of enhancement strategies to close these gaps \cite{yao2025mcqg, dong2025effiqa,wen2025ai,zhao2024retrieval}, but these techniques have not been systematically organized or compared by failure mode. In this paper, we synthesize the literature into three representative enhancement families and instantiate each with a concrete method. The three enhancement methods we study are not merely abstract categories: in this work, they are concretely instantiated, tuned, and integrated by us into practical pipelines for LLM-based development. At the end, we provide structured guidance on \textbf{when} each family helps, \textbf{which} failure patterns it mitigates, and \textbf{where} gaps remain. 

The main contribution of this work is a failure-aware enhancement framework for LLM-based code generation that goes beyond standard progressive prompting. We design and implement three concrete enhancement pipelines—Self-Critique, Multi-Model Collaboration, and Retrieval-Augmented Generation (RAG) assisted, and then derive a taxonomy that links failure types to the strengths and limitations of each pipeline. Our study yields a failure-driven decision framework with quantified time–efficiency trade-offs, giving practitioners actionable guidance on when and how to apply each enhancement method. In what follows, we present the related work in Section II, the methodology in Section III, and the empirical evaluations in Section IV. Section V discusses our decision framework, Section VI outlines the threats to validity, and finally, Section VII concludes the paper.

\section{Background and Related Work}
% Large language models have demonstrated significant promise for automated code generation, with recent work exploring advanced prompting strategies including chain-of-thought reasoning\cite{wei2022chain}, multi-step decomposition\cite{khot2022decomposed,zhou2022least}, and progressive refinement \cite{madaan2023self}. Various enhancement approaches have been proposed to address generation failures, including retrieval-augmented generation providing external documentation and code examples\cite{lewis2020retrieval}, self-critique and iterative refinement methods \cite{chen2023teaching}, and multi-agent collaboration combining complementary models \cite{hong2023metagpt,qian2023communicative}. However, prior work focuses on proposing individual methods without systematic comparison across failure types or practical guidance on method selection for specific project characteristics. Our work addresses this gap through empirical evaluation of three enhancement methods across diverse failure categories, providing a decision framework that maps failure characteristics to appropriate enhancement strategies based on effectiveness and efficiency trade-offs.

Large language models have demonstrated remarkable capabilities in automated code generation, leveraging advanced reasoning techniques to translate natural language requirements into functional implementations. Chain-of-thought (CoT) prompting, introduced by Wei et al.\cite{wei2022chain}, decomposes complex reasoning into intermediate steps, significantly improving LLM performance on mathematical and logical tasks. Building upon CoT, Wang et al. proposed self-consistency reasoning paths achieving substantial accuracy gains on arithmetic benchmarks \cite{wang2022self}. Progressive prompting extends these ideas to software development by structuring code generation through sequential phases: requirements analysis, architectural design, and implementation, which is analogous to traditional software engineering workflows \cite{wei2024requirements}.

% Despite these advances, systematic failures persist. Chen et al.'s evaluation of Codex, which powers GitHub Copilot, demonstrated that on the HumanEval benchmark, the model only achieves 28.8\% correctness in single-shot evaluation. 

Researchers have proposed complementary approaches: RAG augments LLMs with external documentation through differentiable retrieval mechanisms~\cite{lewis2020retrieval}, while Tree-of-Thoughts (ToT), proposed by Yao et al. \cite{yao2023tree}, enables deliberate exploration of multiple solution paths via search algorithms.Yao et al's work on multiple choice question generation, with iterative self-critique, correction, and comparison feedback, demonstrates how LLMs can improve outputs through self-review cycles \cite{dahiya2024leveraging,yao2025mcqg}. Multi-model collaboration strategies  \cite{dong2025effiqa,wen2025ai}, such as those studied by Dong et al. on efficient question-answering with strategic multi-model collaboration on knowledge graphs, leverage complementary strengths of different models for better performance \cite{dong2025effiqa}. Recent surveys document the rapid evolution of LLM-based code generation agents that incorporate planning, iterative refinement, and multi-model collaboration \cite{ding2024reasoning}. However, practitioners lack systematic guidance on which enhancement method to apply for specific failure types, motivating our empirical investigation of failure-aware selection strategies.

\section{Methodology}

\begin{table*}[!t]
\caption{Challenge Project Characteristics}
\label{prject challenge}
\centering
\scriptsize
\begin{tabularx}{\textwidth}{|c|l|l|c|Y|l|}
\hline
\textbf{ID} & \textbf{Project} & \textbf{Languages} & \textbf{Completion} &
\textbf{Missing / incomplete functionality} & \textbf{Failure Type} \\
\hline
P2  & Train Ticket Reservation    & Java/HTML/CSS                 & 16/17 (94.1\%) &
Missing admin profile update endpoint & Local Logic Failures \\
\hline
P8  & Food Delivery Microservices & Java/Shell                    & 23/24 (95.8\%) &
Incomplete RabbitMQ message queue integration & External Integration Failures \\
\hline
P10 & Shopping Cart               & Java/HTML/CSS                 & 20/22 (90.9\%) &
Missing inventory management and cart quantity logic & Local Logic Failures \\
\hline
P14 & Book Store DevOps           & Python/HTML/CSS/HCL/JS        & 29/31 (93.5\%) &
Incomplete Gunicorn multi-process metrics and VM provisioning workflow & Infrastructure Configuration Failures \\
\hline
P18 & Pharmacy Management         & Python/HTML/CSS/JS            & 10/11 (90.9\%) &
Missing patient medication self-management with healthcare business rules & Domain Knowledge Failures \\
\hline
P25 & Expense Tracker             & JS/TypeScript                 & 12/13 (92.3\%) &
Incomplete multi-currency API integration and i18n framework setup & External Integration Failures \\
\hline
\end{tabularx}
\\[-1.5em]
\end{table*}

We conducted a two-phase empirical study to evaluate enhancement strategies for LLM-based code generation. Phase 1 established a baseline by evaluating progressive prompting on 25 open-source projects. Phase 2 investigated three enhancement strategies on 6 challenge projects where baseline progressive prompting failed to achieve complete requirement coverage.

\subsection{Phase 1: Baseline Establishment}

\textbf{Project Selection.} We selected 25 open-source GitHub projects using three criteria: (1) detailed step-by-step implementation instructions in README files, (2) clear project goals and requirements, and (3) numbered feature lists enabling objective task completion evaluation. Projects spanned diverse domains including e-commerce (online bookstores, shopping carts), booking systems (train tickets, hotels, car rentals), healthcare (pharmacy management), microservices (food delivery), search engines (Wikipedia IR), and DevOps infrastructure (containerized deployments). Programming languages included Java, Python, JavaScript, PHP, TypeScript, and Go, with project complexity ranging from 3 to 31 tasks (average: 13 tasks).

\textbf{Evaluation Approach.} We use original GitHub implementation serving as ground truth for requirement coverage. For each project, we compared two code-generation approaches using a state-of-the-art LLM (GPT-5) \cite{OpenAI_GPT-5_2025}: (a) Direct prompting: Code generation directly from requirements without intermediate design phases. Requirements were provided to the LLM with the prompt \textit{``Generate complete code for this project.''} Multiple prompts were sometimes needed due to output length limitations, but no structured workflow (e.g., design or planning phases) was employed, and (b) Progressive prompting: Multi-step approach following the sequence~\cite{wei2024requirements}: \textit{requirements analysis $\rightarrow$ architectural design $\rightarrow$ test case specification $\rightarrow$ code implementation.}

% All baseline experiments used GPT-5.0 (ChatGPT)\cite{OpenAI_GPT-5_2025} as the code generation model. Task completion was evaluated by comparing generated code against requirement specifications extracted from project documentation.

% \textbf{Baseline Results.} Progressive prompting achieved 96.9\% average task completion across 25 projects, significantly outperforming direct prompting (80.5\%) and matching human-level performance (95.7\%). However, 8 projects failed to reach 100\% completion, with failures concentrated in five categories: CRUD operations (admin management), external service integration (message queues, APIs), business logic (both generic and domain-specific), and DevOps infrastructure. These failure cases motivated our Phase 2 investigation of enhancement strategies.

\subsection{Phase 2: Enhancement Strategy Evaluation}

\subsubsection{Challenge Project Selection}
% From the 8 incomplete projects in Phase 1, we selected 6 projects representing diverse failure types and complexity levels:
We focused on the projects where progressive prompting failed to complete all tasks. From the 8 incomplete projects, we selected 6 representative projects following a maximum diversity sampling strategy~\cite{franco2005sampling}. The two excluded projects exhibited redundant failure patterns already covered by our selected samples, shown in the table \ref{prject challenge}. These 6 projects provide comprehensive coverage of observed failure patterns while maintaining experimental feasibility (6 projects $\times$ 3 methods = 18 experiments). 
For each project, we applied three enhancement strategies:

\begin{itemize}[leftmargin=1em, itemsep=1pt, topsep=2pt, parsep=0pt]
    \item \textbf{Self-Critique: } This method applies iterative self-review to baseline progressive prompting. After initial code generation, we prompted the LLM to critique its own output: \textit{``Review the generated code carefully. Is it complete? What requirements are missing or only partially implemented?''} The LLM then identified gaps and generated fixes for incomplete implementations. This critique-fix cycle continued until the LLM reported all requirements complete or no further improvements were identified. Self-Critique uses the same model as baseline progressive prompting, adding only review iterations. The hypothesis is that explicit code review can identify incomplete implementations that baseline prompting missed.
    \item \textbf{Multi-Model Collaboration: } This method combines two different LLMs with complementary strengths: GPT-5 for requirements analysis and architectural design, and Claude Sonnet 4.5 for code implementation. The process has two phases: (1) \textbf{Design phase:} GPT-5 analyzes requirements and produces detailed architectural specifications including component structure, API design, data models, and integration patterns; (2) \textbf{Implementation phase:} The architectural specification is provided to Claude Sonnet 4.5, which generates code following the design. This separation leverages GPT-5's advanced reasoning capabilities for system architecture while utilizing Claude's code generation strengths. The hypothesis is that explicit architectural design reduces ambiguity that causes implementation gaps.
    \item \textbf{RAG-Assisted: } This method augments baseline progressive prompting with retrieved documentation and code examples. Using Claude Sonnet 4.5 with API access, we retrieved relevant materials for each project: (1) official framework documentation (e.g., Spring Boot documentation for P2, Django i18n guides for P25, RabbitMQ tutorials for P8), (2) similar open-source projects from GitHub (e.g., healthcare management systems for P18, microservices with message queues for P8), and (3) implementation patterns matching the project's technology stack. Retrieved materials were provided as context during code generation with the prompt: \textit{``Using these examples and documentation as reference, implement the following requirements...''} The hypothesis is that concrete examples and authoritative documentation reduce implementation ambiguity and provide proven patterns for complex features.
\end{itemize}

\subsubsection{Evaluation Metrics and Process}

For each project-method combination (18 total experiments), we measured five metrics: \textit{task completion rate, improvement over baseline, total LLM generation time, prompt count, and an efficiency metric (minutes per percentage-point (pp) improvement)}. The min/pp metric enables fair comparison across projects with different baseline completion rates and allows practitioners to assess cost-benefit trade-offs.

We assessed task completion through systematic manual evaluation. For each project, we compared generated code against requirement specifications extracted from GitHub README files. Each task was classified as: \textbf{Complete:} Fully implemented with all specified functionality; \textbf{Partial:} Implementation present but missing required features or edge cases; \textbf{Missing:} No implementation found in generated code. 

To analyze enhancement methods from a \emph{failure-aware} perspective, we classify each incomplete task not by low-level bug patterns~\cite{song2023empirical}, but by the kind of \emph{missing functionality} in the target system. In our experiments, models failed less on basic syntax, type correctness, or straightforward control flow; most generated code compiled and followed common framework idioms without manual repair. Instead, the remaining errors clustered around high-level functional omissions~\cite{liu2023your}, along four dimensions: 
\begin{itemize}
    \item Local Logic Failures: Issues solvable through code review (CRUD, validation rules)
    \item External Integration Failures: Requiring API docs, service configs
    \item Domain Knowledge Failures: Requiring specialized expertise
    \item Infrastructure Configuration Failures: Requiring deployment knowledge.
\end{itemize}
They correspond to missing pieces of an application’s functionality, not to specific coding mistakes. In that sense they are a high‑level view of what’s “incomplete” in each project. 

Classification was based on code inspection without execution testing, as projects require complex environment setup (databases, message queues, cloud resources, API keys). To ensure objectivity, we created an evaluation checklist for each project listing all required tasks and their acceptance criteria before examining generated code.

Time measurements captured wall-clock time from first prompt submission to final code generation for each method, excluding time spent by the human evaluator formulating prompts or reviewing intermediate outputs. For Self-Critique, time included all critique-fix cycles. For Multi-Model, time included both GPT-5 design phase and Claude implementation phase. For RAG-Assisted, time included retrieval plus generation with retrieved context. All measurements represent LLM generation time only, not human thinking time.

\section{EVALUATION}

% To analyze enhancement methods from a \emph{failure-aware} perspective, we classify each incomplete task not by low-level bug patterns~\cite{song2023empirical}, but by the kind of \emph{missing functionality} in the target system.  
% From the 8 incomplete tasks left by baseline progressive prompting, we derived six recurring failure types:

% instead group functional omissions. They correspond to missing pieces of an application’s functionality, not to specific coding mistakes. In that sense they are a high‑level view of what’s “incomplete” in each project. 

% This section provides a comprehensive evaluation of baseline performance compared to the effectiveness of our enhancements. Additionally, we discuss the method’s cost-effectiveness and provide a friendly, easy-to-use decision framework for practitioners. We organize our analysis around three research questions: RQ1: How does progressive prompting compare with direct prompting in completing tasks? RQ2: How well do self-critique, multi-model collaboration, and RAG address progressive prompting failures, and how does their effectiveness differ by failure type? RQ3: What time efficiency trade-offs exist among these enhancement methods? All our experimental materials are publicly available at https:[add zenodo link in here].
This section evaluates baseline performance and our enhancements under three research questions.
\textbf{RQ1}: How does progressive prompting compare to direct prompting in task completion? \textbf{RQ2}: How do Self-Critique, Multi-Model Collaboration, and RAG-Assisted perform on progressive prompting failures, and how does their effectiveness vary by failure type? \textbf{RQ3}: What are the time–efficiency trade-offs among these enhancement methods? We also discuss cost-effectiveness and present a practical decision framework for practitioners. Experimental materials are available at https://doi.org/10.5281/zenodo.17637008. 

% RQ1 compares progressive versus direct prompting, RQ2 examines how Self-Critique, Multi-Model, and RAG handle failures by type, and RQ3 analyzes time–efficiency trade-offs among these methods. 

% We also discuss cost-effectiveness and present a practical decision framework for practitioners. Experimental materials are available at [Zenodo link].
% We present our findings in three parts. First, we establish baseline performance showing that progressive prompting significantly outperforms direct prompting but still exhibits failures (RQ1). Second, we evaluate three enhancement strategies on six challenge projects, demonstrating that RAG and Multi-Model achieve near-perfect completion while Self-Critique shows mixed effectiveness (RQ2). Third, we analyze cost-benefit trade-offs, revealing that RAG achieves the best efficiency at 31.5 minutes per percentage point improvement (RQ3).

\subsection{Baseline Performance (RQ1)}

Baseline results (Table~\ref{tab:baseline}) on 25 projects show progressive prompting averaged 96.9\% completion (SD 5.8\%) versus 80.5\% (SD 13.0\%) for direct prompting (Cohen’s $d$ = 1.63, $p < 0.001$). SD denotes the standard deviation of project-level completion rates, and Cohen’s $d$ is a standardized effect size where values above 0.8 are typically interpreted as large. Progressive prompting completed all tasks in 17/25 projects (68\%) compared to 4/25 (16\%) for direct prompting. Human-written solutions averaged 95.7\% completion, indicating that progressive prompting nearly matches human coverage. These findings confirm that progressive prompting significantly outperforms direct prompting (RQ1).

\begin{table}[!t]
\caption{Baseline Performance Comparison (25 Projects)}
\label{tab:baseline}
\centering
\scriptsize
\setlength{\tabcolsep}{3pt} % tighten column spacing
\resizebox{\columnwidth}{!}{%
\begin{tabular}{lcccccc}
\toprule
\textbf{Method} & \textbf{Avg} & \textbf{Median} & \textbf{SD} & \textbf{100\%} & \textbf{$\geq$90\%} & \textbf{Avg} \\
 & \textbf{Compl} & & & \textbf{Rate} & \textbf{Rate} & \textbf{Prompts} \\
\midrule
Human       & 95.7\%         & 100\%          & --          & 16/25 (64\%)           & 22/25 (88\%)           & -- \\
Direct      & 80.5\%         & 80\%           & 13.0\%      & 4/25 (16\%)            & 6/25 (24\%)            & 1.8 \\
Progressive & \textbf{96.9\%} & \textbf{100\%} & \textbf{5.8\%} & \textbf{17/25 (68\%)} & \textbf{24/25 (96\%)} & 20.4 \\
\bottomrule
\end{tabular}%
}
\\[-2em]
% \\[0.25em]
% \footnotesize Statistical test: Wilcoxon signed-rank, $W = 236.5$, $p < 0.001$; Cohen's $d = 1.63$.\\
% \footnotesize \textsuperscript{a}Not applicable (single version, no prompts used).
\end{table}

% Despite progressive prompting's strong performance, 8 projects failed to achieve 100\% completion with completion rates ranging from 90.9\% to 95.8\%. Analysis of the 19 incomplete tasks across these projects revealed five distinct failure categories: CRUD operations (1 project, missing admin endpoints), external service integration (2 projects, incomplete API and message queue setup), business logic errors (2 projects, both generic and domain-specific), and infrastructure configuration (1 project, incomplete observability). These failures demonstrate that progressive prompting alone is insufficient for complete requirement coverage, motivating our evaluation of enhancement strategies.

However, 8 of the 25 projects remained incomplete (completion 90.9\%–95.8\%), comprising 19 missing tasks in four failure categories: Local Logic Failures (1 project), External Integration Failures (2 projects), Domain Knowledge Failures (2 projects), and Infrastructure Configuration Failures (1 project). These gaps show that progressive prompting alone does not achieve full coverage, motivating the need for enhancement strategies.

\subsection{Enhancement Strategy Effectiveness (RQ2)}

% Table~\ref{tab:enhancement_projects} presents completion rates for six challenge projects across three enhancement methods. RAG achieved the highest win rate at 4 of 6 projects (67\%), followed by Self-Critique at 2 of 6 projects (33\%), while Multi-Model achieved zero individual wins despite reaching 100\% completion on 5 of 6 projects. Average completion rates were 96.0\% for Self-Critique (improvement: +3.0pp over baseline), 99.2\% for Multi-Model (+6.3pp), and 99.2\% for RAG (+6.3pp). Both Multi-Model and RAG reached 100\% completion on 5 of 6 projects (83\%), compared to only 2 of 6 for Self-Critique (33\%).

Enhancement results (Table~\ref{tab:enhancement_projects}) on six challenge projects show RAG-Assisted achieved the best performance 4/6 projects (67\%) and Self-Critique 2/6 (33\%), while Multi-Model not rank first in any project even though it reached 100\% completion on 5/6 projects (83\%). Average completion rates were 96.0\% (Self-Critique, +3.0pp), 99.2\% (Multi-Model, +6.3pp), and 99.2\% (RAG-Assisted, +6.3pp).

Failure-type effects: Self-Critique solved P2 (logic failures) and P10 (domain Knowledge failures) with 100\% completion, but made no improvement on P8, P18, and P25 (external/domain/integration failures). RAG-Assisted and Multi-Model succeeded on all projects except P10, where both stalled at 95.5\% due to complex inventory edge cases requiring further refinement.

Integration tasks: On P8, P14, P18, and P25 (external APIs, infrastructure, and domain-specific logic), RAG-Assisted consistently reached 100\% completion while Self-Critique made no progress. This highlights that external documentation and concrete examples are critical for resolving these failure types.

Code-reviewable failures: Projects P2 and P10 had logic or missing-function errors visible in the code. Self-Critique’s iterative review identified these gaps and fixed them, indicating it is most effective on failures that are “code-reviewable” without external knowledge.

Multi-Model and RAG-Assisted both averaged 99.2\% completion, but assisted won 4/6 projects because it completed them faster with higher confidence. Multi-Model’s zero led in none reflect high reliability (5/6 projects at 100\%) but longer runtimes, making it a secondary choice when RAG-Assisted also succeeds. Self-Critique’s 33\% win rate (2/6) highlights its niche strength on code-reviewable failures and its limitations on integration/domain tasks.

% A Friedman test comparing the three methods showed no statistically significant differences ($\chi^2$ = 3.60, df = 2, $p$ = 0.165), though large effect sizes (Cohen's d $>$ 1.0 for all methods vs baseline) suggest practical importance.

A Friedman test \cite{zimmerman1993relative} found no significant difference across methods ($\chi^2$ = 3.60, df = 2, $p$ = 0.165), but large effect sizes (Cohen's d $>$ 1.0 for all methods vs baseline) imply practical differences. Overall, these results (RQ2) confirm that failure type, rather than method sophistication, determines effectiveness.

\begin{table*}[!t]
\caption{Enhancement Results for Six Challenge Projects}
\label{tab:enhancement_projects}
\centering
\scriptsize
\begin{tabular}{llccccccc}
\toprule
\textbf{ID} & \textbf{Project} & \textbf{Tasks} & \textbf{Baseline} & \textbf{Self-C} & \textbf{Multi-M} & \textbf{RAG-A} & \textbf{Best} & \textbf{Improv} \\
\midrule
P2 & Train Ticket & 17 & 94.1\% & \textbf{100\%} & \textbf{100\%} & \textbf{100\%} & All three & +5.9pp \\
P8 & Food Delivery & 24 & 95.8\% & 95.8\% & \textbf{100\%} & \textbf{100\%} & \textbf{RAG-A} (fastest) & +4.2pp \\
P10 & Shopping Cart & 22 & 90.9\% & \textbf{100\%} & 95.5\% & 95.5\% & \textbf{Self-C} & +9.1pp \\
P14 & Book Store & 31 & 93.5\% & 96.8\% & \textbf{100\%} & \textbf{100\%} & \textbf{RAG-A} (fastest) & +6.5pp \\
P18 & Pharmacy & 11 & 90.9\% & 90.9\% & \textbf{100\%} & \textbf{100\%} & \textbf{RAG-A} (fastest) & +9.1pp \\
P25 & Expense Tracker & 13 & 92.3\% & 92.3\% & \textbf{100\%} & \textbf{100\%} & \textbf{RAG-A} (fastest) & +7.7pp \\
\midrule
\multicolumn{2}{l}{\textbf{Average}} & \textbf{19.7} & \textbf{92.9\%} & \textbf{96.0\%} & \textbf{99.2\%} & \textbf{99.2\%} & -- & \textbf{7.1pp} \\
\bottomrule
% \multicolumn{9}{l}{\footnotesize $\checkmark$ indicates method selected as ``best'' based on efficiency or confidence when multiple achieved 100\%}
\end{tabular}
\\[-2.5em]
\end{table*}

\begin{table}[!t]
\caption{Summary Statistics Across Enhancement Methods}
\label{tab:summary_stats}
\centering
\scriptsize
\begin{tabularx}{\columnwidth}{l *{3}{>{\centering\arraybackslash}X}}
\toprule
\textbf{Metric} & \textbf{Self-C} & \textbf{Multi-M} & \textbf{RAG-A} \\
\midrule
Avg Completion      & 96.0\%            & 99.2\%            & 99.2\%            \\
Improvement         & +3.0pp            & +6.3pp            & +6.3pp            \\
100\% Projects      & 2/6 (33\%)        & 5/6 (83\%)        & 5/6 (83\%)        \\
Method rank first         & 2/6 (33\%)        & 0/6 (0\%)         & \textbf{4/6 (67\%)} \\
Avg Time            & \textbf{138 min}  & 279 min           & 193 min           \\
Avg Prompts         & \textbf{23}       & 30                & 27                \\
Efficiency (min/pp) & 40.6*             & 47.4              & \textbf{31.5}     \\
\bottomrule
\multicolumn{4}{l}{\footnotesize *Excludes $\infty$ values for P8, P18, P25 (zero improvement).}
\end{tabularx}
\\[-2em]
\end{table}

% While Multi-Model and RAG achieved identical average completion rates (99.2\%), RAG won 4 of 6 individual project comparisons because it reached 100\% completion faster and with higher confidence estimates. Multi-Model's zero wins reflect its consistent reliability (5/6 projects at 100\%) but higher time cost, making it second choice when both succeed. Self-Critique's 33\% win rate (2/6 projects) demonstrates niche effectiveness for code-reviewable failures but fundamental limitations for integration and domain-specific tasks. A Friedman test comparing the three methods showed no statistically significant differences ($\chi^2$ = 3.60, df = 2, $p$ = 0.165), though large effect sizes (Cohen's d $>$ 1.0 for all methods vs baseline) suggest practical importance. These results address RQ2, revealing that failure type determines method effectiveness.

\subsection{Cost-Benefit Analysis and Efficiency (RQ3)}

% Figure~\ref{fig:efficiency} presents the relationship between improvement (in percentage points) and time required (in minutes) for all 18 experiments. RAG achieved the best average efficiency at 31.5 minutes per percentage point improvement, compared to 47.4 min/pp for Multi-Model and 40.6 min/pp for Self-Critique (excluding failures where Self-Critique provided zero improvement). Time efficiency differed significantly across methods (Friedman test: $\chi^2$ = 12.0, df = 2, $p$ = 0.0025), with all pairwise comparisons statistically significant ($p$ = 0.031). The efficiency advantage stems from RAG's ability to directly apply proven patterns from retrieved examples, reducing trial-and-error iterations compared to Multi-Model's architecture-first approach or Self-Critique's iterative refinement.

\begin{figure}[!t]
\centering
\includegraphics[width=0.48\textwidth]{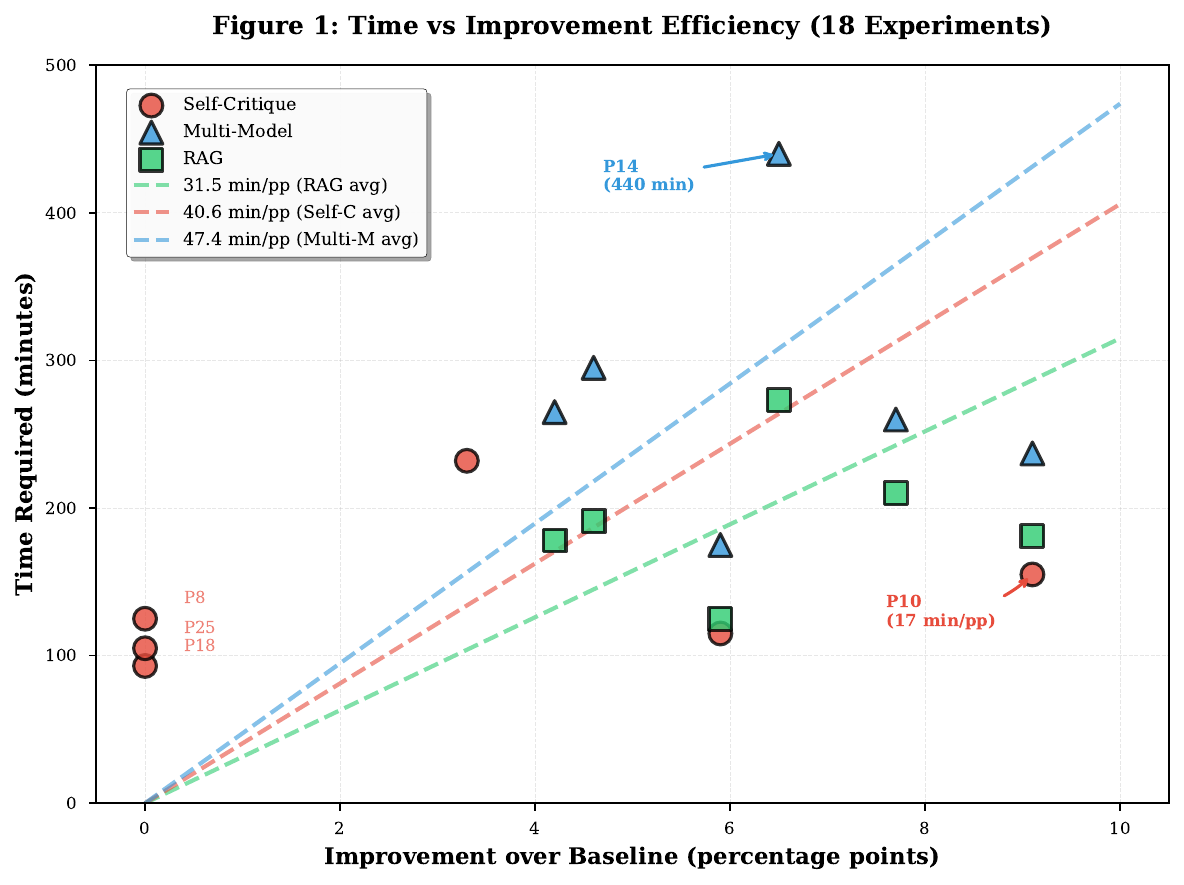}
\\[-1em]
\caption{Time vs Improvement Efficiency. RAG-Assisted (green squares); Self-Critique (red circles); Multi-Model (blue triangles)}
\label{fig:efficiency}
\vspace{-1em}
\end{figure}

Figure~\ref{fig:efficiency} shows the improvement vs. time for all 18 experiments. RAG-Assisted was most efficient (31.5 min per percentage point), compared to 40.6 for Self-Critique and 47.4 for Multi-Model. Efficiency differed significantly (Friedman test: $\chi^2$ = 12.0, df = 2, $p$ = 0.0025; all pairwise $p$ = 0.031). RAG-Assisted’s advantage comes from directly applying retrieved patterns, avoiding the trial-and-error of Multi-Model’s design first or Self-Critique’s iterative approaches. These efficiency results highlight three practical considerations when choosing 
an enhancement method:

Speed vs. Reliability: Self-Critique is fastest when it works (138 min avg) but fails on non-code-reviewable tasks. RAG-Assisted balances speed and reliability (193 min avg) across all tasks. Multi-Model is slowest (279 min) but most reliable (5/6 projects 100\% completion), making it the choice when correctness is paramount.

Prompts vs. Quality: Multi-Model uses the most prompts (30 on average), RAG-Assisted about 27 (including retrieval), and Self-Critique 23. However, prompt count did not reflect solution quality: Self-Critique’s fewer prompts stem from its failure on complex tasks, not from efficiency.

Upfront Cost vs. Adaptation: RAG-Assisted needs an initial setup (API access, documentation) but then adapts efficiently to different failure types. Multi-Model needs no special setup but consistently incurs high time cost. Self-Critique requires no setup but can’t handle failures that aren’t visible in code.

Overall, RAG-Assisted offers the best cost–benefit (31.5 min/pp) while maintaining high reliability (99.2\% completion), addressing RQ3.

\section{Discussion}

\subsection{Enhancement}

Our results reveal that enhancement method effectiveness depends critically on failure characteristics rather than method sophistication. 

% The five failure types identified in baseline progressive prompting---CRUD operations, external service integration, generic business logic, domain-specific business logic, and infrastructure configuration---exhibit distinct patterns that explain why Self-Critique, Multi-Model, and RAG succeed or fail.

\textbf{Self-Critique Failed on External Service Integration.} P8 (Food Delivery) required RabbitMQ integration. Self-Critique reviewed code but couldn't identify missing queue configurations—these aren't inferrable from application code alone. RAG-Assisted retrieved Spring Cloud RabbitMQ tutorials with complete patterns (exchanges, routing keys, listeners), enabling direct implementation. This explains RAG-Assisted's 100\% success on all external integration projects (P8, P14, P18, P25) versus Self-Critique's zero improvement.

\textbf{Self-Critique Succeeded on Business Logic.} Project P10 required inventory management validation logic, specifically preventing checkout when requested quantity exceeds available stock. Self-Critique's code review identified this gap and the LLM's critique noted ``The checkout endpoint does not verify that cart.quantity $\leq$ product.stock before processing payment,'' and subsequent iterations added the missing validation. This success demonstrates Self-Critique's strength: when failures stem from incomplete or incorrect logic that is reviewable through code inspection, iterative refinement effectively identifies and fixes gaps.
Notably, Multi-Model and RAG-Assisted also got 95.5\% on P10, missing the inventory check. GPT-5’s design covered basic cart flow but omitted validation rules, and RAG-Assisted’s retrieved examples lacked inventory logic (often assuming unlimited stock). This indicates RAG-Assisted’s limitation: it depends on having relevant examples; without them, it cannot fill the gap.

\textbf{RAG-Assisted Achieves Superior Efficiency.} RAG-Assisted’s 31.5 min/pp efficiency (vs 40.6 and 47.4) comes from directly using retrieved patterns. For example, RAG-Assisted fetched Django i18n setup for P25, enabling immediate correct code. In contrast, Multi-Model adds design overhead and Self-Critique needs many review cycles or fails on missing external knowledge. Figure~\ref{fig:efficiency}’s results confirm a bimodal outcome: Self-Critique either completes quickly (e.g. P2 in 115 min, P10 in 155 min) or yields no improvement (P8, P18, P25). Code-reviewable failures (logic failures) are solved fastest by Self-Critique, while failures needing external knowledge cannot be resolved by Self-Critique regardless of effort.

\subsection{Decision Framework for Practitioners}

Our results show that Self-Critique, Multi-Model Collaboration, and Retrieval-Augmented Generation do not compete as mutually exclusive alternatives; instead, they attack complementary aspects of progressive prompting failures. When we apply all three enhancements, almost all remaining tasks can be recovered. However, in practice, running every enhancement on every failure is rarely desirable due to latency and cost. To make our findings actionable, we compare the methods along two axes: complete rates by failure type and time cost, and distill them into a hierarchical decision framework summarized in Table~\ref{tab:decision}.

The framework follows a simple principle: for each failure type, start with the fastest method that actually works reliably on that type, and escalate only when needed. For local logic failures that are visible in the code (CRUD operations, missing functions, validation rules), Self-Critique is both the fastest and the most effective. In contrast, Self-Critique shows no improvement on external integration, domain-knowledge, and infrastructure failures; in these cases, RAG-Assisted becomes the first-line choice. Multi-Model Collaboration is slower than either Self-Critique or RAG-Assisted but provides the highest reliability overall, so we position it as a second option: used when correctness is paramount or when faster methods leave residual gaps. This hierarchical view is a trade-off solution derived from our empirical data.

Practitioners can plug in their own cost constraints and failure profiles, and still reuse the same ordering principle. This decision rules can be encoded directly as a meta-prompt for LLM-based agents: given a partially implemented project and a failing task, the agent (1) classifies the failure into one of our types, (2) selects the corresponding first-line enhancement, and (3) escalates to the second and third methods only if the failure persists. In this sense, our framework not only explains when each enhancement helps, but also serves as a reusable prompt structure for orchestrating LLM tools in practice.

\begin{table}[!t]
\caption{Hierarchical decision framework by failure type}
\label{tab:decision}
\centering
\scriptsize
\begingroup
\setlength{\tabcolsep}{2pt} % 默认大约是 6pt，这里改小一点
\begin{tabular}{@{}lccc@{}}
\toprule
\textbf{Failure Type} & \textbf{1st Choice} & \textbf{2nd Choice} & \textbf{3rd Choice} \\
\midrule
Local Logic Failures & Self-Critique & RAG-A & Multi-M \\
External Integration Failures & RAG-A & Multi-M & Self-Critique \\
Domain Knowledge Failures & RAG-A & Multi-M & Self-Critique \\
Infrastructure Configuration Failures & RAG-A & Multi-M & Self-Critique \\
\bottomrule
\end{tabular}
\\[-2em]
\endgroup
\end{table}

\section{Threats to Validity}
\subsection{Internal Validity}

% Time measurements depend on hardware configuration (CPU, GPU, RAM), network latency, and API server load, introducing variability in absolute timing values. We conducted all experiments on a single machine with consistent network conditions, ensuring valid relative comparisons across methods. While absolute times may differ on other hardware, we expect method rankings (RAG most efficient, Multi-Model slowest) to remain stable, as they reflect inherent algorithmic differences rather than hardware-specific optimizations.

% Task completion assessment was conducted manually, introducing potential subjectivity in classifying implementations as complete, partial, or missing. We mitigated this through pre-defined evaluation checklists specifying required functionality for each task, reducing interpretation variability. However, edge cases between ``complete'' and ``partial'' implementation required judgment calls.

% Our study evaluates specific LLM implementations (GPT-5 Thinking Mode, Claude Sonnet 4.5), and results may vary with different models or future versions. However, our decision framework focuses on enhancement method characteristics (iterative refinement, architectural separation, retrieval augmentation) rather than model-specific features, suggesting broader applicability across LLM platforms.

We manually judged task completion (complete/partial/missing), which introduces subjectivity. To mitigate this, we used predefined checklists of required functionality, reducing variability. Still, some edge cases between “complete” and “partial” required judgment. Our study evaluates specific LLM implementations (GPT-5, Claude Sonnet 4.5), and results may vary with different models or future versions. However, our decision framework focuses on enhancement method characteristics (iterative refinement, architectural separation, retrieval augmentation) rather than model-specific features, suggesting broader applicability across LLM platforms.

\subsection{External Validity}

Our 25 GitHub projects span a range of domains and complexity levels (3–31 tasks) and were intentionally selected to have detailed implementation instructions in their README files. This design choice improves reproducibility and allows us to measure requirement coverage in a controlled way, but it may also bias our sample toward relatively well-documented systems compared to projects with vaguer specifications.

The projects cover multiple programming languages and mainstream application domains. We therefore expect our failure taxonomy and decision framework to be most representative of common web and enterprise development settings. At the same time, the results may not fully generalize to highly specialized domains (e.g., embedded systems, scientific computing, safety-critical systems). Extending and validating the framework in these specialized contexts remains an important direction for future work.

% Our 25 GitHub projects, while spanning diverse domains and complexity levels (3-31 tasks), may not represent all software development contexts. Projects were selected for detailed implementation instructions in README files, potentially favoring well-documented systems over typical projects with vaguer specifications.

% Validation on proprietary codebases and projects with less detailed requirements would strengthen generalizability.

% While our projects span multiple programming languages (Java, Python, JavaScript, PHP, TypeScript, Go) and domains (e-commerce, healthcare, microservices, infrastructure), results may not generalize to specialized domains (embedded systems, scientific computing, real-time systems) or languages with different characteristics (functional languages, low-level systems programming). The failure taxonomy and decision framework should apply to mainstream web and enterprise application development.

% Time efficiency measured LLM generation time but excluded human effort in prompt formulation, code review, and post-generation debugging. Total development time would include human time, though automation of repetitive generation tasks still provides value. Our measurements enable method comparison under consistent conditions, with relative efficiency differences reflecting inherent method characteristics.

Finally, our time-efficiency results are influenced by factors such as
hardware configuration, network connectivity, and API load, so the
reported wall-clock times should be interpreted as context-dependent
rather than universal constants. In line with prior work on productivity
and efficiency metrics in software engineering, time-based measures as \emph{relative} indicators to compare alternatives in a
given context, not as precise predictions of absolute performance \cite{sadowski2019rethinking}. In this sense, our failure-aware decision framework is designed as a simple and actionable heuristic that captures the three patterns and offers practitioners an intuitive, friendly first-step guideline for choosing an appropriate enhancement strategy.

\section{Conclusion and Future Work}
% Our empirical study answers three research questions about LLM code generation enhancement. We demonstrate that progressive prompting significantly outperforms direct code generation (RQ1: 96.9\% vs 80.5\%, Cohen's d=1.63, $p<0.001$) but exhibits systematic failures requiring enhancement strategies. We reveal that enhancement method effectiveness depends critically on failure characteristics (RQ2): Self-Critique succeeds on code-reviewable logic errors but fails completely on external service integration, while RAG achieves consistent success across all failure types with superior efficiency. We show significant time efficiency differences (RQ3: $p=0.0025$), with RAG achieving optimal cost-benefit trade-off at 31.5 minutes per percentage point improvement. Most importantly, we provide a decision framework mapping failure types to appropriate enhancement methods, enabling practitioners to select strategies based on project characteristics rather than trial-and-error experimentation.

Our study shows that progressive prompting markedly improves requirement coverage (96.9\% vs 80.5\%) but still leaves some tasks unimplemented. Enhancement effectiveness varies by failure type: iterative self-critique corrects logical errors but fails on external integrations, while RAG-Assisted completes all failure types with higher efficiency. RAG-Assisted also achieved the best time–efficiency (about 31.5 minutes per percentage-point improvement). We propose a failure-driven decision framework that links each failure pattern to the most effective enhancement strategy, enabling practitioners to select methods based on project characteristics rather than trial-and-error. Our contributions include three concrete enhancement pipelines, a taxonomy of failure modes, and evidence-based guidelines for AI-assisted code generation.
In future work, we plan to automate failure-type diagnosis and enhancement selection, for example by learning classifiers or policies that can be embedded into LLM-based development agents. We also plan to validate and refine the decision framework on larger, industrial codebases.
% Future work should explore automated failure type classification to enable adaptive enhancement selection without manual diagnosis. Hybrid approaches combining Self-Critique's iterative refinement with RAG's external knowledge retrieval could address both logic errors and integration gaps efficiently, providing practitioners with even more effective code generation enhancement strategies.

\bibliographystyle{ieeetr}
\bibliography{bibi}

@article{wei2022chain,
  title={Chain-of-thought prompting elicits reasoning in large language models},
  author={Wei, Jason and Wang, Xuezhi and Schuurmans, Dale and Bosma, Maarten and Xia, Fei and Chi, Ed and Le, Quoc V and Zhou, Denny and others},
  journal={Advances in neural information processing systems},
  volume={35},
  pages={24824--24837},
  year={2022}
}

@article{wang2022self,
  title={Self-consistency improves chain of thought reasoning in language models},
  author={Wang, Xuezhi and Wei, Jason and Schuurmans, Dale and Le, Quoc and Chi, Ed and Narang, Sharan and Chowdhery, Aakanksha and Zhou, Denny},
  journal={arXiv preprint arXiv:2203.11171},
  year={2022}
}

@article{lewis2020retrieval,
  title={Retrieval-augmented generation for knowledge-intensive nlp tasks},
  author={Lewis, Patrick and Perez, Ethan and Piktus, Aleksandra and Petroni, Fabio and Karpukhin, Vladimir and Goyal, Naman and K{\"u}ttler, Heinrich and Lewis, Mike and Yih, Wen-tau and Rockt{\"a}schel, Tim and others},
  journal={Advances in neural information processing systems},
  volume={33},
  pages={9459--9474},
  year={2020}
}

@inproceedings{ding2024reasoning,
  title={Reasoning and planning with large language models in code development},
  author={Ding, Hao and Fan, Ziwei and Guehring, Ingo and Gupta, Gaurav and Ha, Wooseok and Huan, Jun and Liu, Linbo and Omidvar-Tehrani, Behrooz and Wang, Shiqi and Zhou, Hao},
  booktitle={Proceedings of the 30th ACM SIGKDD Conference on Knowledge Discovery and Data Mining},
  pages={6480--6490},
  year={2024}
}

@article{yao2023tree,
  title={Tree of thoughts: Deliberate problem solving with large language models},
  author={Yao, Shunyu and Yu, Dian and Zhao, Jeffrey and Shafran, Izhak and Griffiths, Tom and Cao, Yuan and Narasimhan, Karthik},
  journal={Advances in neural information processing systems},
  volume={36},
  pages={11809--11822},
  year={2023}
}

@inproceedings{wei2024requirements,
  title={Requirements are all you need: From requirements to code with llms},
  author={Wei, Bingyang},
  booktitle={2024 IEEE 32nd International Requirements Engineering Conference (RE)},
  pages={416--422},
  year={2024},
  organization={IEEE}
}

@inproceedings{dong2025effiqa,
  title={Effiqa: Efficient question-answering with strategic multi-model collaboration on knowledge graphs},
  author={Dong, Zixuan and Peng, Baoyun and Wang, Yufei and Fu, Jia and Wang, Xiaodong and Zhou, Xin and Shan, Yongxue and Zhu, Kangchen and Chen, Weiguo},
  booktitle={Proceedings of the 31st International Conference on Computational Linguistics},
  pages={7180--7194},
  year={2025}
}

@inproceedings{yao2025mcqg,
  title={Mcqg-srefine: Multiple choice question generation and evaluation with iterative self-critique, correction, and comparison feedback},
  author={Yao, Zonghai and Parashar, Aditya and Zhou, Huixue and Jang, Won Seok and Ouyang, Feiyun and Yang, Zhichao and Yu, Hong},
  booktitle={Proceedings of the 2025 Conference of the Nations of the Americas Chapter of the Association for Computational Linguistics: Human Language Technologies (Volume 1: Long Papers)},
  pages={10728--10777},
  year={2025}
}

@inproceedings{song2023empirical,
  title={An empirical study of code generation errors made by large language models},
  author={Song, Da and Zhou, Zijie and Wang, Zhijie and Huang, Yuheng and Chen, Shengmai and Kou, Bonan and Ma, Lei and Zhang, Tianyi},
  booktitle={7th Annual Symposium on Machine Programming},
  year={2023}
}

@article{zhao2024retrieval,
  title={Retrieval-augmented generation for ai-generated content: A survey},
  author={Zhao, Penghao and Zhang, Hailin and Yu, Qinhan and Wang, Zhengren and Geng, Yunteng and Fu, Fangcheng and Yang, Ling and Zhang, Wentao and Jiang, Jie and Cui, Bin},
  journal={arXiv preprint arXiv:2402.19473},
  year={2024}
}

@inproceedings{wen2025ai,
  title={The AI Imitation Game: A Cognitive Comparison of Mimicry in Large Language Models},
  author={Wen, Victor and Peng, Zedong and Chen, Yusi},
  booktitle={2025 IEEE International Conference on Information Reuse and Integration and Data Science (IRI)},
  pages={79--84},
  year={2025},
  organization={IEEE}
}

@misc{OpenAI_GPT-5_2025,
  author = {OpenAI},
  title = {Response from GPT-5 (Model: GPT-5-Turbo, Example Version)},
  howpublished = {Accessed: 14 Nov. 2025},
  year = {2025},
  note = {Prompt used: "Explain how to cite large language models in BibTeX." (See Appendix A for full transcript)},
  url = {https://chat.openai.com/}
}

@article{franco2005sampling,
  title={A sampling strategy for conserving genetic diversity when forming core subsets},
  author={Franco, Jorge and Crossa, Jos{\'e} and Taba, Suketoshi and Shands, Henry},
  journal={Crop Science},
  volume={45},
  number={3},
  pages={1035--1044},
  year={2005},
  publisher={Wiley Online Library}
}

@article{liu2023your,
  title={Is your code generated by chatgpt really correct? rigorous evaluation of large language models for code generation},
  author={Liu, Jiawei and Xia, Chunqiu Steven and Wang, Yuyao and Zhang, Lingming},
  journal={Advances in Neural Information Processing Systems},
  volume={36},
  pages={21558--21572},
  year={2023}
}

@article{zimmerman1993relative,
  title={Relative power of the Wilcoxon test, the Friedman test, and repeated-measures ANOVA on ranks},
  author={Zimmerman, Donald W and Zumbo, Bruno D},
  journal={The Journal of Experimental Education},
  volume={62},
  number={1},
  pages={75--86},
  year={1993},
  publisher={Taylor \& Francis}
}

@book{sadowski2019rethinking,
  title={Rethinking productivity in software engineering},
  author={Sadowski, Caitlin and Zimmermann, Thomas},
  year={2019},
  publisher={Springer Nature}
}

@inproceedings{dahiya2024leveraging,
  title={Leveraging chatgpt to predict requirements testability with differential in-context learning},
  author={Dahiya, Mahima and Gill, Rashminder and Niu, Nan and Gudaparthi, Hemanth and Peng, Zedong},
  booktitle={2024 IEEE International Conference on Information Reuse and Integration for Data Science (IRI)},
  pages={170--175},
  year={2024},
  organization={IEEE}
}
\end{document}